\documentclass[prl,twocolumn,amsmath,amssymb,amsfonts,superscriptaddress,floatfix,showpacs,aps]{revtex4-1}
\newcommand{\beq}{\begin{equation}}
\newcommand{\eeq}{\end{equation}}
\usepackage{graphics}
\usepackage{amsmath}
\usepackage{amsfonts}
\usepackage{amssymb}
\usepackage{graphicx}
\usepackage{color}
\usepackage[normalem]{ulem}
\usepackage{notes2bib}

\begin{document}

\title{Fluence dependent delay of Ni in an FeNi alloy supports an exchange based origin}

\author{Somnath Jana}
\email[Email: ] {somnath.jana@mbi-berlin.de/sj.phys@gmail.com}
 \affiliation{Max Born Institute for Nonlinear Optics and Short Pulse Spectroscopy, Max-Born-Str. 2a, 12489 Berlin, German}
\author{Ronny Knut}
 \affiliation{Department of Physics and Astronomy, Uppsala University, Box 516, 75120 Uppsala, Sweden}
 \author{Puloma Singh}
 \affiliation{Max Born Institute for Nonlinear Optics and Short Pulse Spectroscopy, Max-Born-Str. 2a, 12489 Berlin, German}
 \author{Kelvin Yao}
 \affiliation{Max Born Institute for Nonlinear Optics and Short Pulse Spectroscopy, Max-Born-Str. 2a, 12489 Berlin, German}
\author{Christian Tzschaschel}
 \affiliation{Max Born Institute for Nonlinear Optics and Short Pulse Spectroscopy, Max-Born-Str. 2a, 12489 Berlin, German}
\author{Johanna Richter}
 \affiliation{Max Born Institute for Nonlinear Optics and Short Pulse Spectroscopy, Max-Born-Str. 2a, 12489 Berlin, German}
\author{Daniel Schick}
 \affiliation{Max Born Institute for Nonlinear Optics and Short Pulse Spectroscopy, Max-Born-Str. 2a, 12489 Berlin, German}
\author{Denny Sommer}
 \affiliation{Max Born Institute for Nonlinear Optics and Short Pulse Spectroscopy, Max-Born-Str. 2a, 12489 Berlin, German}
\author{Dieter Engel}
 \affiliation{Max Born Institute for Nonlinear Optics and Short Pulse Spectroscopy, Max-Born-Str. 2a, 12489 Berlin, German}
 \author{Olof Karis}
 \affiliation{Department of Physics and Astronomy, Uppsala University, Box 516, 75120 Uppsala, Sweden}
\author{Clemens von Korff Schmising}
 \affiliation{Max Born Institute for Nonlinear Optics and Short Pulse Spectroscopy, Max-Born-Str. 2a, 12489 Berlin, German}
\author{Stefan Eisebitt}
 \affiliation{Max Born Institute for Nonlinear Optics and Short Pulse Spectroscopy, Max-Born-Str. 2a, 12489 Berlin, German}
 \affiliation{Technische Universität Berlin, Straße des 17. Juni 135, 10623 Berlin}

\begin{abstract}
The delayed demagnetization in Ni relative to Fe in the ultrafast demagnetization studies in FeNi alloy has led to two competing theoretical explanations: The Inhomogeneous Magnon Generation (IMG) and the Optically Induced Spin Transfer (OISTR) model. The IMG attributes the delay to the preferential magnon generation at the Fe sites and its subsequent propagation to Ni, while OISTR proposes direct spin transfer from Ni to Fe. In this study, we employ element-resolved extreme ultraviolet spectroscopy to investigate the effect of excitation strength on this delay, aiming to resolve the controversy. The data indicate a significant reduction in the delay with increasing fluence, which is inconsistent with the theoretical predictions of OISTR. These findings, in conjunction with the observation of a saturation of Fe demagnetization at the onset of Ni demagnetization, indicate that a spin-wave instability within the IMG framework may provide a potential explanation for the experimental results.
\end{abstract}

\maketitle

\section{Introduction}
Despite decades of experimental and theoretical efforts, the microscopic process that governs the ultrafast demagnetization in ferromagnetic metals has remained highly debated ~\cite{Carva2013, Koopmans2010, Battiato2012}. 
While separating the various competing processes involving photons, electrons, spins and phonons is challenging, advancement of measuring techniques such as element specific spectroscopy in the extreme ultraviolet (XUV) spectral range have enabled the tracking of elemental magnetization with few tens of femtosecond temporal resolution~\cite{Chan_2009,Jana2017,Yao2020}. 
Minute differences in the transient magnetization behavior between different atomic constitutes of an alloy could be revealed and new hypotheses were proposed~\cite{Mathias2012, Turgut2013, Guenther_2015FeRh, knut2018inhomogeneous, Hofherr_2020_OISTR, Willems_2020_OISTR, Gupta_2023}.
An intriguing observation in this context is the delay of the Ni demagnetization transient relative to that of the Fe in an FeNi alloy~\cite{Mathias2012,Jana2017,JanaFeNi_slicing2022}. 
This was originally reported by Mathias et al. ~\cite{Mathias2012}, who proposed an exchange driven origin for the delay.
Subsequently, two theoretical models have been proposed to explain this delay: Knut et. al. put forth an inhomogeneous magnon generation (IMG) model~\cite{knut2018inhomogeneous}, while Hofherr et. al. attribute the delay to OISTR~\cite{Hofherr_2020_OISTR}. 

IMG considers a preferential ultrafast magnon generation at the Fe sites, relying on a much larger $s$-$d$ exchange interaction strength compared to the Ni sites. 
In a magnon emission process via $s$-$d$ exchange, an up-spin at the localized 3$d$ band is exchanged with a down-spin at the itinerant band. Magnon emission at the Fe site thus reduces the local Fe moment, while the up-spin electron scattered into the itinerant band finally loses angular momentum to the lattice by SOC mediated spin-flips. The bottleneck of magnetization loss is therefore set by the electron-phonon spin-flip rate. 
Magnon emission at the Ni site is hindered by the low $s$-$d$ exchange rate. This is aggravated by the fact that the itinerant band is continuously being filled by the up-spin electrons via the efficient magnon emission process at the Fe sites. Once the magnon created at the Fe sites propagates to the Ni sites via direct exchange, demagnetization of the local Ni moment sets in. The delay is therefore a measure of the magnon propagation time.

Hofherr et al. proposed a direct, optically induced spin moment transfer between the Fe and Ni~\cite{Hofherr_2020_OISTR}. Time dependent density functional theory (TDDFT) predicts a direct transfer of minority electrons from Ni site to the Fe site, attributed to the large density of unoccupied minority states at the Fe site. This results in demagnetization in Fe and an increase in the Ni magnetic moment. 
The increase of the Ni moment is experimentally correlated with an increase in magnetic asymmetry at 2 eV below the Ni $M$-edge, as measured in a transverse magneto-optical Kerr effect (T-MOKE) geometry. 
The observed delay in the demagnetization measured at Ni $M$-edges is consequently rationalized by two competing processes: OISTR driven increase and SOC-mediated demagnetization driven decrease together manifesting as a delayed response. 

Meanwhile, a simulation based on the TDDFT predicts a dramatic increase of the OISTR effect upon increasing excitation pulse strength (fluence), resulting in its complete dominance over the SOC mediated spin-flip process~\cite{Elhanoty_2022_FePd}. The simulation was performed for a FePd alloy, a conceptually similar system, with more unoccupied minority Fe states giving rise to OISTR from the occupied Pd minority states.

In this study, we have investigated the impact of fluence on the delay of Ni demagnetization relative to Fe in an FeNi alloy, seeking to resolve the aforementioned controversy. Contrary to the prediction of the OISTR mechanism, our results reveal a decreasing delay in Ni demagnetization relative to Fe with increasing fluence. Supported by the observed saturation of Fe demagnetization at the onset of Ni demagnetization, we propose that a spin-wave instability, within the framework of the IMG model, may underlie the reduced delay of the Ni demagnetization with increasing fluence.

\begin{figure}[t]
\includegraphics[width=0.99\columnwidth]{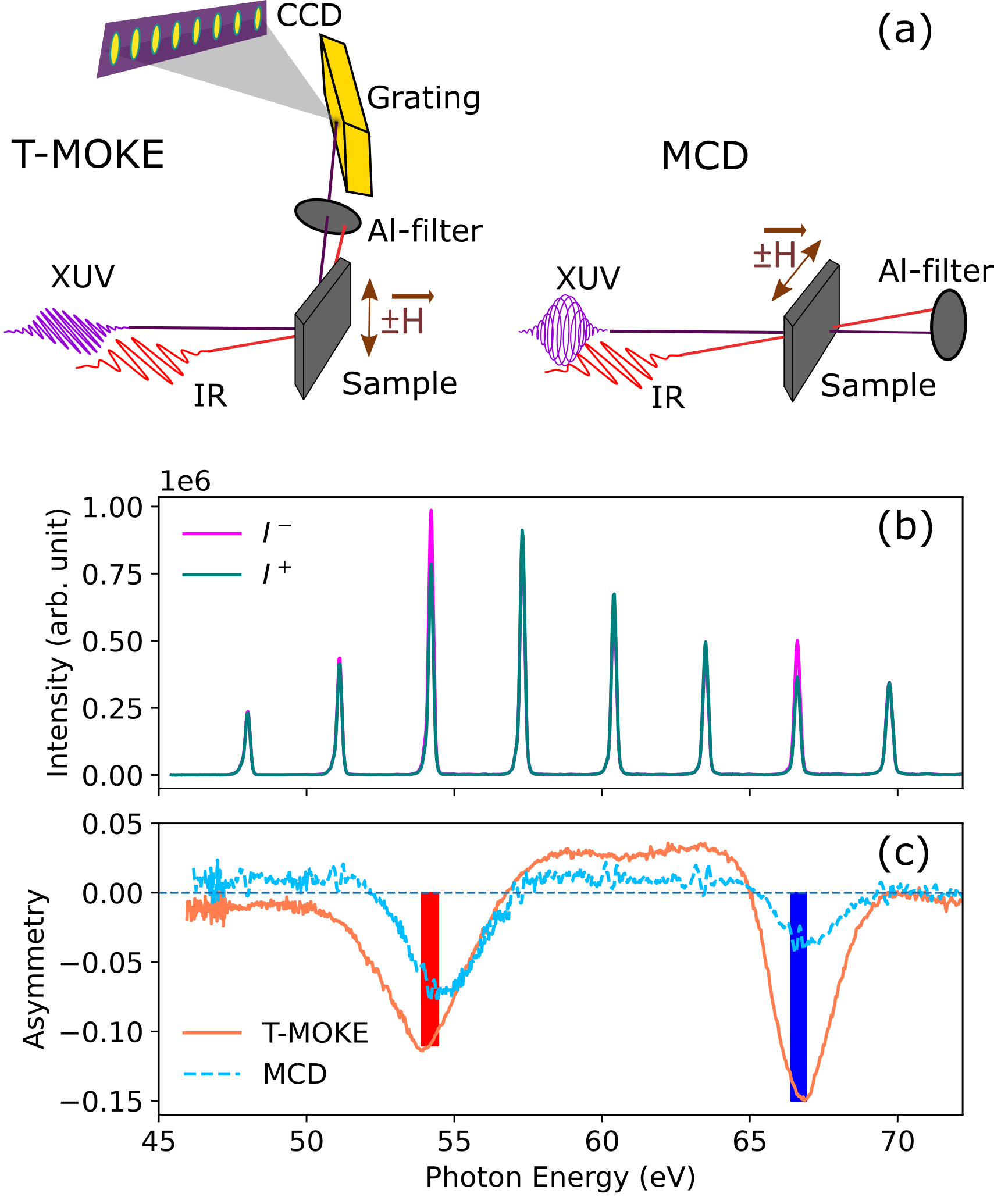}
\caption{(Color Online) (a) displays schematics of the measurement geometries for T-MOKE (left) and MCD (right). (b) shows typical XUV spectra collected for two opposite magnetization directions and (c) shows the asymmetry spectra obtained for T-MOKE and MCD in transmission geometries in the unpumped state of a Fe$_{0.5}$Ni$_{0.5}$ sample. The red and the blue shaded regions in (c) mark the energy ranges at the Fe and Ni M-edges where time-resolved data are collected.\label{Fig_Spectrum}}
\end{figure}

\section{Experimental methods}
Ta(3)/Fe$_{0.5}$Ni$_{0.5}$(25)/Ta(3)/Si-substrate and Ta(3)/ Fe$_{0.5}$Ni$_{0.5}$(25)/Ta(3)/Si(100)/Al(100) samples were prepared by electron-beam evaporation for measurements in T-MOKE and transmission geometries, respectively. The numbers in parentheses indicate the layer thicknesses in nanometers. The thin Si membrane facilitates XUV transmission in the transmission geometry, while a 100 nm Al layer deposited on the back side enhances heat dissipation.

The experimental setup utilizes high harmonic generation (HHG) from a femtosecond near-infrared (NIR) laser with a central wavelength of 800\,nm, 25\,fs pulse width, 3\,mJ pulse energy, at 3\,kHz repetition rate~\cite{Yao2020}. About 90\% of the linearly polarized NIR pulse is focused into a He-filled gas cell to produce ultrashort XUV pulses that retain the polarization and coherence of the driving NIR \cite{Corkum1993}. The (as generated) linearly polarized XUV pulses are employed for measuring the transverse magneto-optical Kerr (T-MOKE) signal in the reflection geometry [c.f. left inset of Fig. \ref{Fig_Spectrum} (a)]. Furthermore, a four-mirror polarizer is utilized to convert the XUV radiation to circularly polarized XUV radiation~\cite{Schmising_circularXUV_2017, Willems_2015}, enabling the measurement of magnetic circular dichroism (MCD) in transmission geometry. In both geometries, the incidence angle on to the sample was set to 45$^{\circ}$ and a grating, placed behind the sample, spectrally resolved the XUV radiation on a in-vacuum CCD camera (c.f. Fig.\ \ref{Fig_Spectrum}(a)). A similar grating and CCD camera combination is used upstream to measure a reference spectrum in order to reduce noise due to XUV source intensity fluctuations. The magnetization is probed by calculating the asymmetry parameter, $A(E) = [I^+(E) - I^- (E)]/[I^+(E) + I^- (E)]$, where $I^{\pm}(E)$ is the reflected/transmitted intensity measured for opposite magnetic field directions ($\pm$$H$), applied parallel to the sample surface. 
The magnetic field is applied perpendicular (parallel) to the plane of incidence in case of measurements in the T-MOKE (transmission) geometry (c.f. Fig. \ref{Fig_Spectrum} (a)). The use of transitions involving core orbitals in both the T-MOKE and MCD signals allows for a large atomic specificity in the magnetization transients.

A portion of the NIR ($\sim$ 10\%) is directed to the pump line, which consists of a delay stage, a combination of a $\lambda/2$ wave-plate and a reflective polarizer to set the excitation fluence, 
and a telescope to control the size of the pump spot to a full width at half maximum area of $\sim$ 0.3 mm$^2$. The fluences reported in this paper are incident fluences. Based on the nominal sample structure and using methods described in Ref. [\onlinecite{Khorsand_2014}], we estimate about 16\% of the NIR gets absorbed in the sample (see Supplementary Material (SM)). For further details of the experimental setup, refer to Ref. [\onlinecite{Yao2020}].

\begin{figure*}[hbt!]
\includegraphics[width=1.99\columnwidth]
{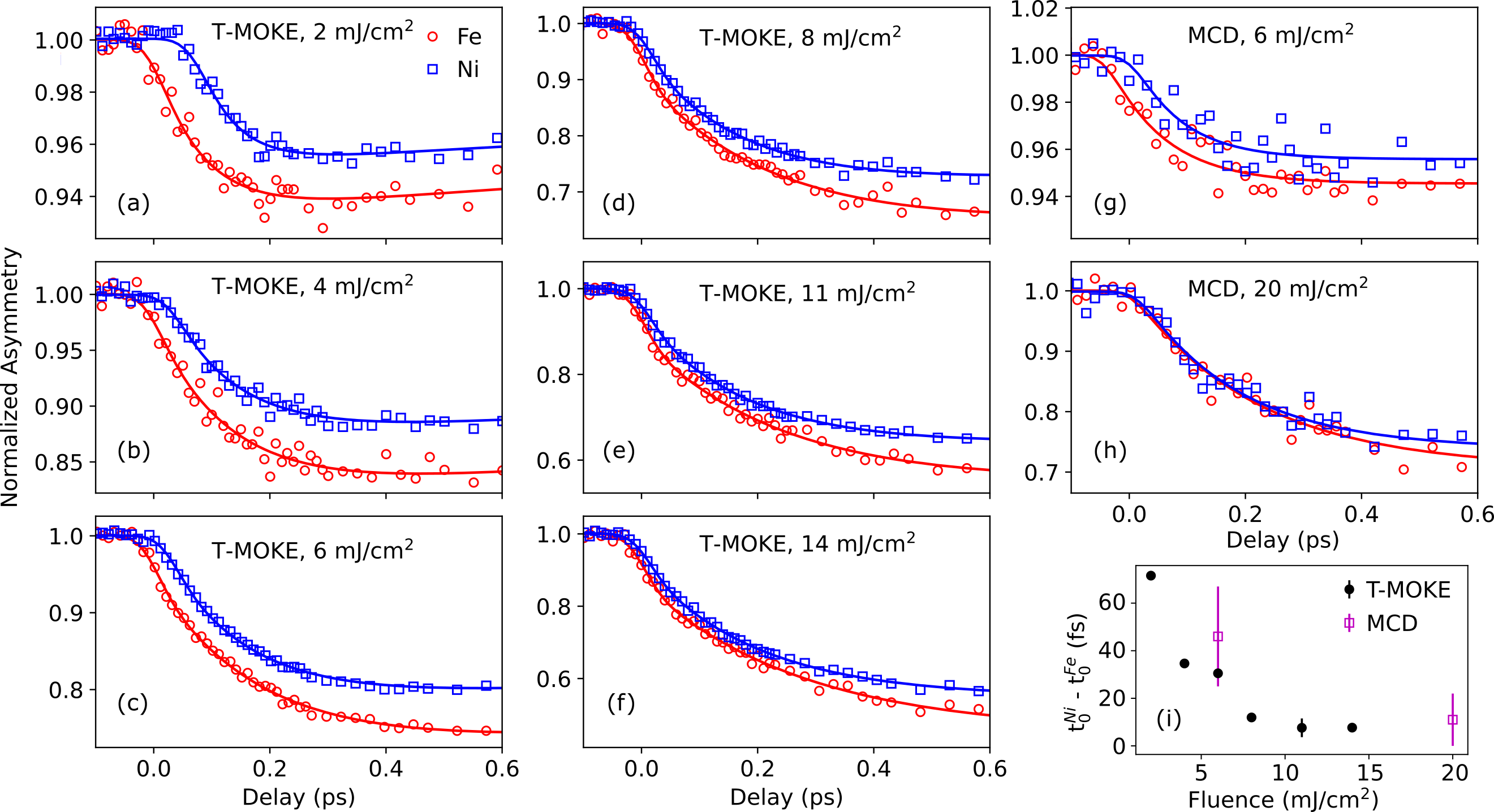}
\caption{(Colour Online) (a) - (h) show normalized T-MOKE/MCD asymmetries vs pump-probe delay measured at the Fe and Ni M-edges in Fe$_{0.5}$Ni$_{0.5}$. (i) shows delay in the onset of Ni demagnetization compared to that of Fe as a function of fluence. \label{Fig2}}
\end{figure*}

\section{Results and discussions}
Typical XUV spectra in reflection collected for opposite magnetization direction are shown in Fig. \ref{Fig_Spectrum}(b). The photon energies of the harmonics are approximately 3.1\,eV separated as only the odd harmonics of the NIR are generated. Continuous asymmetry spectra, collected in both the T-MOKE and MCD geometries, are displayed in Fig. \ref{Fig_Spectrum}(c). These continuous spectra result from the combination of multiple data sets collected by adjusting the NIR pulse width, pulse energy and He gas pressure in the gas cell in the HHG generation process aimed at shifting the harmonics and thus effectively generating photons across a range of energies. The apparent difference in spectral shape between the T-MOKE and MCD asymmetries arises from the fact that, while both the dispersive ($\delta$) and absorptive ($\beta$) components of the refractive index ($n = 1 - \delta + \mathrm{i} \beta$) contribute to the T-MOKE asymmetry obtained in reflection, only the absorptive component affects the MCD measured in transmission. Fortuitously, the harmonics near the Fe ($\sim$ 54.2\,eV) and Ni ($\sim$ 66.7\,eV) M-edges align close to the peaks of the asymmetries in both geometries, as indicated by the red and blue shaded regions in Fig. \ref{Fig_Spectrum}(c) (also compare with Fig. \ref{Fig_Spectrum}(b)). This alignment facilitates the measurement of element-specific demagnetization dynamics using both geometries. Due to low transmission and smaller asymmetry, the data measured in the transmission geometry exhibits a lower signal to noise ratio compared to the T-MOKE measurements. Although T-MOKE asymmetry can exhibit nonlinearities under certain conditions~\cite{Hennecke_depthProfile, Richter_PRB_2024}, particularly at energies where the asymmetry crosses zero, the energy positions used in this study (indicated by color bars in Fig. \ref{Fig_Spectrum}(c)) and the observed asymmetry values satisfy the conditions for a linear relationship. Furthermore, simulations of the T-MOKE asymmetry confirm the absence of nonlinearity with magnetization, as well as from nonuniform demagnetization along the depth, at both the Fe and Ni M-edges for the studied sample (see the SM for details). Additionally, the MCD signal remains unaffected by potential nonlinearity issues, making it a valuable cross-check.

Time-resolved data were collected in the T-MOKE geometry at six different incident fluences: 2, 4, 6, 8, 11, 14\,mJ/cm$^2$. Fig. \ref{Fig2}(a - f) display normalized asymmetries as a function of pump-probe delay at both the Fe and Ni M-edges in Fe$_{0.5}$Ni$_{0.5}$ for all six fluences. 
The scattered data points in the figure represent the measured asymmetries, while the solid lines correspond to fitted curves obtained using a fit function incorporating exponential decay, as described in the SM. Assuming a linear relationship between asymmetry and magnetization, a clear delay in the onset of Ni demagnetization compared to that of the Fe is observed for all six fluences. As shown in Fig. 2(i), the delay, which is as high as 72\,fs at the lowest fluence, decreases with increasing fluence, eventually reaching below 15\,fs at higher fluences. We will discuss the ramifications of this important observation in detail below.

The results of recent measurements conducted using a T-MOKE geometry have demonstrated a pronounced dependence of asymmetry dynamics on the incidence angle in the vicinity of the Ni M-edge in both Ni and FeNi alloys~\cite{Probst_2024, Möller_2024}. These results were rationalized on the basis of a disproportionate change between the dispersive ($\delta$) and absorptive ($\beta$) components of the refractive index, with the effect being more pronounced near the zero crossing of the asymmetry spectra. Since the MCD asymmetry in transmission depends only on $\beta$, dynamics measured in transmission should eliminate the ambiguities with respect to the delays observed in T-MOKE measurements. 
Therefore, pump-probe MCD asymmetries were measured at two fluences: $\sim$ 6 and $\sim$ 20 mJ/cm². The experimental data are presented in Fig. \ref{Fig2}(g) and (h). Despite very small absolute changes of the asymmetry, on the order of only 0.001, a noticeable delay of 42 ± 21 fs can be observed at the lower fluence, which -- at higher fluence -- almost disappears (11 $\pm$ 11 fs). As shown in Fig. \ref{Fig2}(i), the delay as extracted from the transmission data follows a similar fluence-dependent trend as the T-MOKE data within the error margins.

As previously mentioned, two theoretical models have been proposed to explain the distinct elemental dynamics in FeNi. Hofherr et. al. \cite{Hofherr_2020_OISTR} propose OISTR to explain the delay in Ni.
As the mechanism suggests, a signature of OISTR is the \textit{fast and mirrored dynamics}, where one element (Ni) gains spin moment while the other element (Fe) loses spin moment due to intersite spin transfer~\cite{Dewhurst_2018, Elhanoty_2022_FePd}. 
As discussed in the introduction, the delay in Ni demagnetization has been attributed to two competing processes: an OISTR-driven increase and a SOC mediated demagnetization-driven decrease in the transient magnetic signal at the Ni M-edge. 
Further, the effect of OISTR significantly intensifies with higher excitation fluence in Ref. \cite{Elhanoty_2022_FePd}, showcasing its predominant influence over SOC-mediated spin-flip processes during the initial time scales. 
Therefore, with increasing fluence, an increase in the delay of Ni demagnetization -- or even an enhancement of the Ni moment -- is expected. However, our experimental results show a decrease in the delay (c.f. Fig. \ref{Fig2}(i)) with increasing fluence, and no enhancement of the Ni moment is observed. 
Thus, our results contradict the prediction based on OISTR described in Ref. \onlinecite{Elhanoty_2022_FePd}.

Knut et. al. \cite{knut2018inhomogeneous} suggested an explanation for the delay of the Ni demagnetization based on an element dependent ultrafast magnon generation. Upon optical excitation, electrons are excited to the conduction band and undergo scattering at both phonons and magnons. The rate of the demagnetization depends on the electron-phonon spin-flip rate, while the delay of the onset of magnetization decay at Ni sites results from an inhomogeneous magnon generation. Due to the preferential ultrafast magnon generations at the Fe sites, Fe loses spin moment, while the Ni demagnetization only starts when the spin waves reach the Ni sites. The delay in Ni demagnetization is thus a measure of the magnon propagation time, which depends on the spin wave stiffness ($D_\mathrm{spin}$). We note that Ref.~\cite{Jana_2023} employed this IMG model to extract $D_{spin}$ from measured delays of the Ni demagnetization as a function of concentration $x$ in Fe$_{1-x}$Ni$_x$ alloys yielding results in good agreement with values obtained by other experimental techniques as well as by ab-initio theory.

\begin{figure}[t]
\includegraphics[width=0.99\columnwidth]{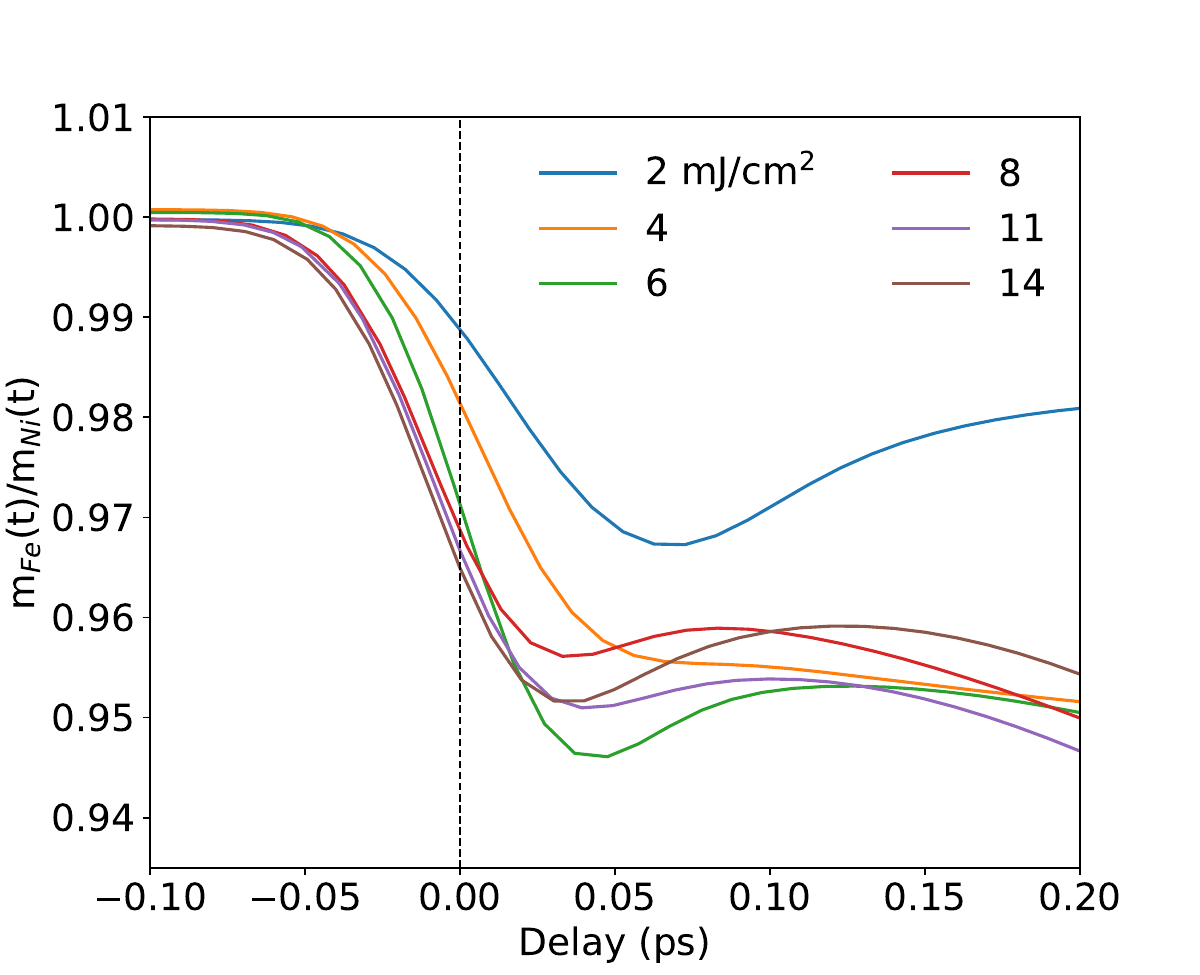}
\caption{(Colour Online)  The ratio of normalized magnetization of Fe and Ni is plotted over the pump-probe delay, highlighting the saturation of magnon generation in Fe. \label{Fig_magnon}}
\end{figure}

Since the delay depends on $D_\mathrm{spin}$ within the IMG model, the observed decrease in delay with increasing fluence needs further explanation. 
One possible explanation is the saturation of the IMG at higher fluence levels, analogous to the saturation effect observed in microwave absorption during ferromagnetic resonance experiments due to Suhl's instability \cite{Suhl_1957}. 
In the IMG framework, ultrafast magnons are predominantly generated at Fe sites and propagate to Ni sites, with the delay reflecting this propagation time. 
To sustain the inhomogeneity of the magnon generation profile as the fluence increases, a higher number of magnons must be generated at the Fe sites within a fixed time window, causing larger tilts of Fe spins relative to Ni spins.  
This, in turn, activates a large exchange energy between Fe and Ni spins, potentially destabilizing the magnon modes at Fe-sites and causing them to break into different spin-wave modes, ultimately disrupting the inhomogeneity of the magnon profile. 
Notably, our data reveal that Fe demagnetization reaches a maximum of about 5\% at the onset of Ni demagnetization. This is illustrated in Fig. \ref{Fig_magnon} by calculating the ratio of the normalized magnetization of Fe relative to Ni plotted against the pump-probe delay.
The curves in Fig. \ref{Fig_magnon}, derived from fitted demagnetization data, exhibit a minimum near the onset of Ni demagnetization. Intriguingly, the Fe demagnetization appears to reach a maximum of 5$\pm$1\% at these minima, suggesting the saturation in the difference of magnons at Fe-sites compared to the Ni-sites.
This observation implies that Fe spins achieve a maximum tilt angle of approximately 18° ($\cos^{-1}(0.95)$) relative to Ni spins before breaking into different spin-wave modes, which also involve spins at Ni sites. The apparent decrease in the delay of Ni demagnetization can therefore be attributed to a spin-wave instability arising from the large tilt angle of the Fe spins as fluence increases.

\begin{figure}[t]
\includegraphics[width=0.99\columnwidth]{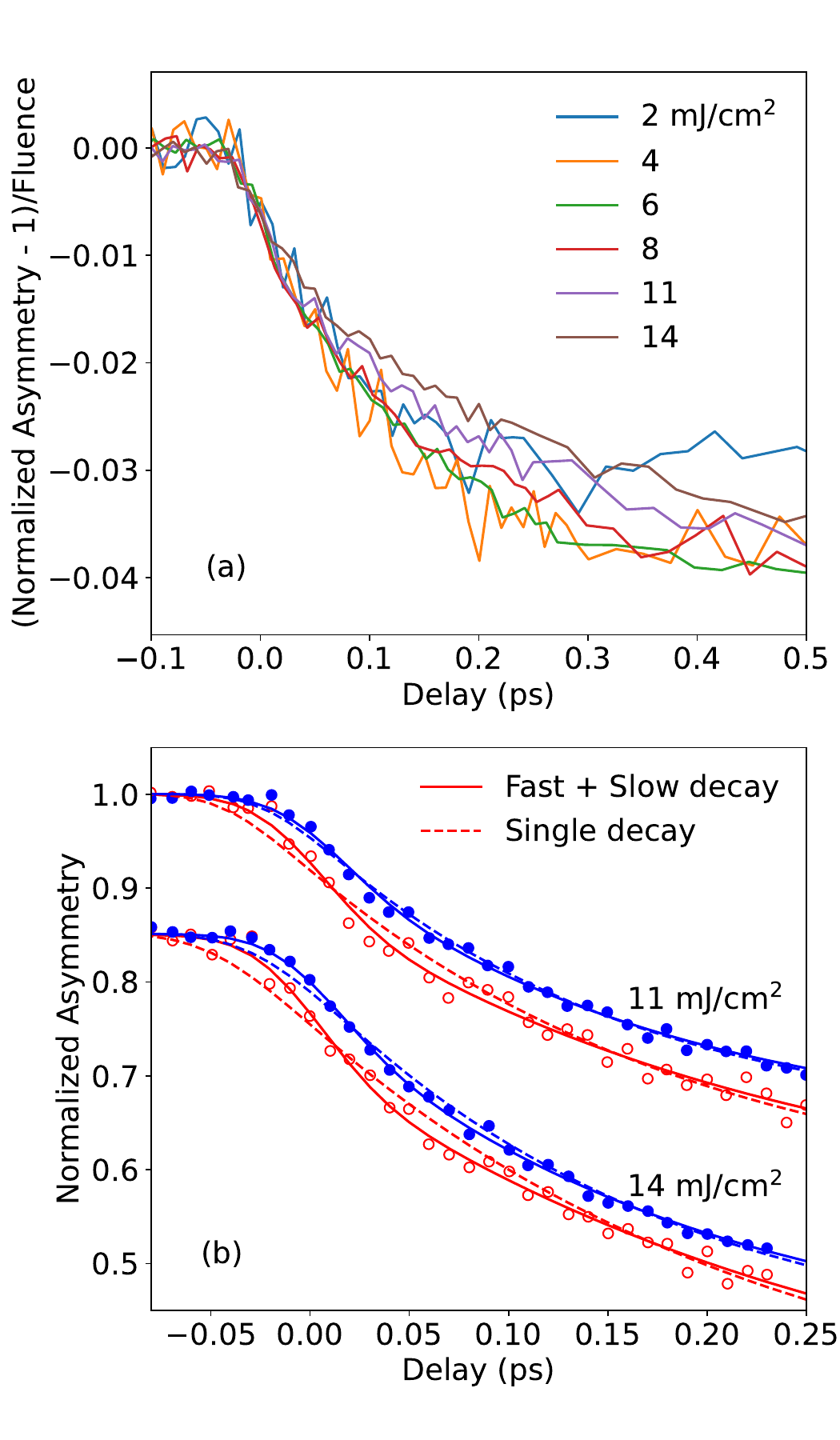}
\caption{(Colour Online) (a) Change of T-MOKE asymmetries vs pump-probe delay measured at the Fe M-edge are plotted after linearly scaling by the incident fluence. (b) Fits of the data measured at 11\,mJ/cm$^2$ and 14\,mJ/cm$^2$ (shifted vertically for clarity) are shown using both single and double exponential decay models. Red:Fe, Blue:Ni. \label{Fig_fastdecay}}
\end{figure}

To further investigate the influence of the excitation fluence, asymmetries at Fe M-edge scaled by their respective fluence values are presented in Fig. \ref{Fig_fastdecay}(a). The Fe asymmetries exhibit an identical decay behavior across all fluences during early times (delay $<$\,100\,fs), indicating that the demagnetization at the Fe site scales with the pump fluence in this time window. This behavior is in contrast to the strong fluence dependence with a pronounces temporal shift observed for Ni demagnetization.  Interestingly, a closer look at the Fe magnetization dynamics reveals a very fast demagnetization during the presence of the pump pulse. This is demonstrated in Fig. \ref{Fig_fastdecay}(b) for the measurements taken at fluences of 11\,mJ/cm$^2$ and 14\,mJ/cm$^2$.
Attempts to fit the normalized asymmetries with a single exponential decay (dashed lines) function did not yield satisfactory results for pump-probe delay $<$\,100\,fs. Fitting the data with a model incorporating two exponential decay constants (solid lines) reveals a very fast decay component during the initial times at both the Fe and Ni edges while providing a satisfactory fit to the experimental data.
The obtained time constant for the fast decay converges to a very small value (see Table 1 in the SM), approaching zero regardless of the fluence level, suggesting the possibility of a distinct decay mechanism concurrent with the pump pulse.
We recall that in the IMG model the bottleneck for demagnetization rate is the electron-phonon spin-flip scattering rate, which -- according to calculations by Carva et al. \cite{Carva2013} -- is significantly enhanced in the presence of a non-equilibrium electron distribution, i.e. in the initial stage after optical excitation. Based on this, we propose that the faster demagnetization observed in the early stages is driven by the presence of a non-equilibrium electron distribution. 

A recent fluence-dependent study on half-metallic
Co$_2$MnGa also reports on a theoretically predicted increase in the OISTR signature with increased fluence~\cite{Ryan_2023}. However, similar to our results, they observed the opposite behavior experimentally. It was speculated that this was due to additional demagnetization contributions that were not included in the theoretical model.

\section{Conclusion}
A fluence-dependent, element-resolved ultrafast demagnetization study is conducted to address the controversy regarding the observed delayed Ni demagnetization as compared to the Fe demagnetization in an FeNi alloy. We observe a \textit{decrease} of the delay in Ni demagnetization with \textit{increasing} fluence, a trend opposite to that predicted by the OISTR mechanism. Instead, within the IMG framework, a saturation of the preferential magnon generation at Fe sites with increasing fluence, attributed to a spin-wave instability at large excitation fluence, could explain the experimental observation. In addition, a fast decay component has been observed on the sub-50 fs timescale, possibly arising from enhanced electron-phonon spin flip scattering rates in the presence of a non-equilibrium electron distribution. A metallic random alloy such as FeNi can be expected to have low potential for exhibiting a measurable OISTR effect, since the lifetimes of excited electronic states should be in the few femtosecond timescale. However, this study does not exclude the potential for observable OISTR in materials that exhibit excitations with longer lifetimes, such as e.g. half metals.

\section{Acknowledgement}
We acknowledge financial support from the Deutsche Forschungsgemeinschaft (DFG, German Research Foundation) – Project-ID 328545488 – TRR 227, project A02. R.K. acknowledge support by the Swedish Research Council
(VR 2021-5395) and the Knut and Alice Wallenberg Foundation (KAW Lightmatter).

\bibliography{FeNi}

\end{document}